\documentclass[showpacs,twocolumn,aps,prl,floatfix,superscriptaddress,10pt]{revtex4-1}
\usepackage{graphicx}
\begin{document}

\title{Broken Symmetries, Zero-Energy Modes, and Quantum Transport in Disordered Graphene: From Supermetallic to Insulating Regimes}

\author{Alessandro Cresti}
\affiliation{IMEP-LAHC (UMR CNRS/INPG/UJF 5130), Grenoble INP, Minatec, 3, Parvis Louis N\'eel, BP 257, F-38016 Grenoble, France}

\author{Frank Ortmann}
\affiliation{CIN2 (ICN-CSIC) and Universitat Aut\'{o}noma de Barcelona, Catalan Institute of Nanotechnology, Campus UAB, 08193 Bellaterra, Spain}

\author{Thibaud Louvet}
\affiliation{CIN2 (ICN-CSIC) and Universitat Aut\'{o}noma de Barcelona, Catalan Institute of Nanotechnology, Campus UAB, 08193 Bellaterra, Spain}
\affiliation{Ecole Normale Superieure de Lyon, 46, All\'{e}e d'Italie, 69007 Lyon France}

\author{Dinh Van Tuan}
\affiliation{CIN2 (ICN-CSIC) and Universitat Aut\'{o}noma de Barcelona, Catalan Institute of Nanotechnology, Campus UAB, 08193 Bellaterra, Spain}

\author{Stephan Roche}
\affiliation{CIN2 (ICN-CSIC) and Universitat Aut\'{o}noma de Barcelona, Catalan Institute of Nanotechnology, Campus UAB, 08193 Bellaterra, Spain}
\affiliation{ICREA, Instituci\'{o} Catalana de Recerca i Estudis Avan\c{c}ats, 08070 Barcelona, Spain}

\begin{abstract}
The role of defect-induced zero-energy modes on charge transport in graphene is investigated using Kubo and Landauer transport calculations. By tuning the density of random distributions of monovacancies either equally populating the two sublattices or exclusively located on a single sublattice, all conduction regimes are covered from direct tunneling through evanescent modes to mesoscopic transport in bulk disordered graphene. Depending on the transport measurement geometry, defect density, and broken sublattice-symmetry, the Dirac point conductivity is either exceptionally robust against disorder ({\it supermetallic state}) or suppressed through a gap opening or by algebraic localization of zero-energy modes, whereas weak localization and the Anderson insulating regime are obtained for higher energies. These findings clarify the contribution of zero-energy modes to transport at the Dirac point, hitherto controversial.
\end{abstract} 

\maketitle

The electronic transport properties of graphene are known to be very peculiar with unprecedented manifestations of quantum phenomena such as Klein tunneling \cite{KAT_NP2,YOU_NP5}, weak antilocalization \cite{MCC_PRL97,TIK_PRL103}, or the anomalous quantum Hall effect \cite{NOV_N438,ZHA_N438}, all driven by a $\pi$-Berry phase stemming from graphene sublattice symmetry and pseudospin degree of freedom \cite{CAS_RMP81,NOV_RMP83,DAS_RMP83}. These fascinating properties, yielding high charge mobility \cite{MOR_PRL100,TAN_PRL99}, are robust as long as disorder preserves a long range character. 
The fundamental nature of transport precisely at the Dirac point is, however, currently a subject of fierce debate and controversies. Indeed, for graphene deposited on oxide substrates, the nature of low-energy transport physics (as its sensitivity to weak disorder) is masked by the formation of electron-hole puddles \cite{DAS_RMP83}. 
A remarkable experiment has, however, recently demonstrated the possibility to screen out these detrimental effects \cite{PON_NP7}, providing access to the zero-energy Dirac physics. 
An unexpectedly large increase of the resistivity at the Dirac point was tentatively related to the Anderson localization \cite{PON_NP7,EVE_RMP80} of an unknown physical origin and questioned interpretation \cite{DAS_PRB85}.

Of paramount importance are therefore the low-energy impurity states known as zero-energy modes (ZEMs) \cite{PER_PRL96,PER_PRB77}, whose impact on the Dirac-point transport physics needs to be clarified. 
ZEMs are predicted or observed for a variety of disorder classes, as topological defects (mainly vacancies) \cite{PER_PRB73,PER_PRB77}, adatoms covalently bonded to carbon atoms \cite{ROB_PRL101,WEH_PRL105}, and extended defects as grain boundaries \cite{VTU_NL13,LAH_NN5}. 
As recently confirmed by scanning tunneling microscopy experiments on graphene monovacancies \cite{UGE_PRL104}, ZEMs manifest as wave functions that decay as the inverse of the distance from the vacancy, exhibiting a puzzling quasilocalized character, whose consequences on quantum transport remain, to date, highly controversial. First, ZEMs have been  predicted to produce a {\it supermetallic regime} by enhancing the Dirac-point conductivity above its minimum ballistic value $\sigma_{\rm min}=4e^{2}/\pi h$ \cite{OST_PRB74,OST_PRL105}, an unprecedented conducting state, which could be, in principle, explored experimentally \cite{TWO_PRL96,KAT_EPJB51,MIA_SCI317}. Second, a similar increase of the Dirac-point conductivity with the defect density has been also reported in the diffusive regime of two-dimensional disordered graphene in the presence of vacancies or adatoms \cite{WEH_PRL105,YUA_PRB82}. These results contrast with the semiclassical conductivity found with the Boltzmann approach \cite{SHO_JPSJ67,PER_PRB73,NOM_PRL98,STA_PRB76,ZHU_PRB85}, and suggest the absence of quantum interferences and localization effects observed for other types of disorder \cite{LHE_PRL106,RAD_PRB86,LIA_PRL109}. Finally, transport experiments in intentionally damaged graphene also report on puzzling conductivity fingerprints, whose physical origin remains to be fully understood \cite{CHE_PRL102,YAN_PRL107}. A comprehensive picture of the role of ZEMs on quantum transport properties in disordered graphene is therefore crucially missing and demands for further theoretical and experimental inspection.  

This Letter provides an extensive analysis of the contribution of zero-energy modes to quantum conduction close to the Dirac point in disordered graphene.  Using the Kubo-Greenwood and Landauer transport approaches, different regimes are numerically explored by changing the aspect ratio of the transport measurement geometry, and by tuning vacancy density and sublattice symmetry breaking features. The robustness of the supermetallic state induced by ZEMs is shown to be restricted to very low densities of compensated vacancies (equally distributed among both sublattices). This occurs as long as tunneling through evanescent modes prevails. In the absence of contact effects, an increase of the conductivity above $4e^{2}/\pi h$ is obtained for the semiclassical conductivity at the Dirac point and ascribed to a high density of ZEMs, but the quantum conductivity analysis unequivocally reveals a localization regime. For a totally uncompensated vacancy distribution (populating a single sublattice), the delocalization of ZEMs in real space is strongly prohibited for a large energy window around the Dirac point owing to the formation of a gap, whereas no appreciable difference of high energy transport (above the gap) is found compared with the compensated vacancy case. 
We would like to mention that some interesting cases of uncompensated impurities and defects have been reported experimentally \cite{ZHO_SCI333,LV_SR2,BAL_NM9}, whose results demand further exploration. 

\textit{System description and methodology}.-
We consider a finite concentration $n$ of vacancies either distributed at random exclusively on one of the two sublattices ($n_{A} = n$, the number of vacancies  per carbon atoms in sublattice A and $n_{B} = 0$, uncompensated case), or equally distributed vacancies on both sublattices ($n_{A}=n_{B} = n/2$, compensated case). The electronic and transport properties are investigated by using a tight-binding model with a single $p_{\rm z}$ orbital {\it per} atom and first nearest neighbor coupling. We model the vacancies by removing the corresponding orbitals from the Hamiltonian \cite{PER_PRL96,PER_PRB77}. To investigate the various transport regimes, two complementary approaches are used.  For studying two-dimensional (bulk) disordered graphene, real-space quantum wave packet dynamics and Kubo conductivity are calculated \cite{LHE_PRL106,ROC_PRL79,ROC_PRB59,ORT_EPL94,RAD_PRB86,ROC_SSC152}. 
The zero-frequency conductivity $\sigma (E,t)$ for energy $E$ and time $t$ is given by $\sigma (E,t)= e^{2} \rho(E) \Delta X^{2}(E,t)/ t$, where $\rho(E)$ is the density of states (DOS) and $\Delta X^{2}(E,t)$ is the mean quadratic displacement of the wave packet at energy $E$ and time $t$:
\begin{equation}
  \Delta X^{2}(E,t) = \frac{\displaystyle {\rm Tr}\bigl[ \delta(E-{\cal H}) | \hat{X}(t)- \hat{X}(0) |^2 \bigr]} 
                           { \strut\displaystyle {\rm Tr}[\delta(E-{\cal H})]} \ . 
  \label{DeltaX2}
\end{equation}
A key quantity in the analysis of the transport properties is the diffusion coefficient $D(E,t)=\Delta X^{2}(E,t)/t$.
In disordered systems, $D(t)$ generally starts with a short-time ballistic motion followed by a saturation regime, which allows us to estimate the transport (elastic) mean-free path $\ell_e$ from its maximum value as $\ell_e(E)=D^{\text{max}}(E)/2 v(E)$ where $v(E)$ is the velocity. The semiclassical conductivity $\sigma_{sc}$ is given analogously by the maximum conductivity. Depending on disorder strength, $D(E,t)$ is found to decay at longer times owing to quantum interferences, whose strength may dictate  weak or strong (Anderson) localization at the considered time scale. Calculations are performed for systems containing several millions of carbon atoms, allowing the capture of all relevant transport regimes. 
\begin{figure}[tpb]
	\begin{center}
	\resizebox{8cm}{!}{\includegraphics{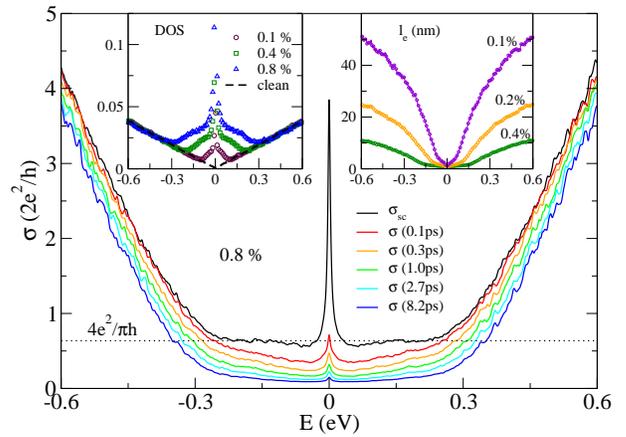}}
	\caption{Main frame: Conductivity of graphene with $n$=0.8\% (compensated case): semiclassical value $\sigma_{sc}$ (solid line), $\sigma_{\rm min}=4e^{2}/\pi h$ (dotted line), and Kubo conductivity at various time scales. Left inset: DOS for varying vacancy density, together with the pristine case (dashed line). Right inset: Mean-free paths for $n$ = 0.1\%, 0.2\%, and 0.4\%.}
  \label{figAB}
  \end{center}
\end{figure}
We also study the ballistic limit of transport through finite graphene samples, by considering strip geometries with width $W$ and length $L$ (with $W/L\gg1$) between two highly doped semi-infinite ribbons (of identical width). This two-terminal transport geometry gives access to the contribution of ZEMs in graphene transport when the charge flow is conveyed by contact-induced evanescent modes. The doping of contacts is simulated by adding an on site energy of -1.5 eV to the corresponding orbitals, which generates a large DOS imbalance between the contacts and the central strip at the Dirac point ($E=0$). The zero-temperature conductivity of the graphene strip is then computed as $\sigma(E)=(2e^2/h)T(E)L/ W$, where $T(E)$ is the transmission coefficient evaluated within the Green's function approach \cite{CRE_PRB76,CRE_NR1}. When $L\ll W$, low-energy transport is dominated by tunneling through the undoped region yielding a universal ballistic value $\sigma(E\approx0)\approx\sigma_{\rm min}= 4e^{2}/\pi h$ at the Dirac point for clean strips \cite{TWO_PRL96,KAT_EPJB51,MIA_SCI317,CRE_PRB76}. 

\textit{ZEMs effects in two-dimensional disordered graphene}.- 
We start by considering the compensated case, which globally preserves the sublattice symmetry. Figure \ref{figAB} (left inset) gives the density of states of the system as a function of the energy $E$ for different vacancy densities $n$.  In agreement with prior results \cite{PER_PRL96,PER_PRB77}, the DOS shows the rise of a broad peak around $E=0$, which witnesses the presence of ZEMs generated by disorder. Their nature, however, is not encoded in this feature but needs to be analyzed by studying transport characteristics such as the mean-free path (Fig.\ref{figAB}, right inset) and conductivity (Fig.\ref{figAB}, main frame). 
The mean-free path $\ell_e$ is seen to be strongly energy dependent with minimum values close to the Dirac point, as expected for short-range scatterers \cite{ORT_EPL94,ROC_SSC152}. By increasing the vacancy density within the range $[0.1\%,0.4\%]$, $\ell_e$ drops from tens of nanometers down to few nanometers, and roughly varies as $\ell_e\sim 1/n$, which is in agreement with the Fermi golden rule. Interestingly, we find for the semiclassical conductivity $\sigma_{sc}\sim E$ for a high enough energy (above 0.3 eV for $n=0.8\%$), whereas it saturates to $\sigma_{\rm min}$ at low energy with a higher value around the Dirac point owing to the DOS enhancement induced by midgap states. When increasing the vacancy density, the minimum conductivity $4e^{2}/\pi h$ around the Dirac point extends over a larger energy region (not shown here). 

The obtained short $\ell_e$ and minimum semiclassical conductivities suggest a strong contribution of quantum interferences, which is further evidenced by the decay of the Kubo conductivity below $\sigma_{\rm min}$ for sufficiently long time scales, see Fig.\ref{figAB} (main frame). Depending on the energy, the observed downscaling of the quantum conductivity versus time can be described by a logarithmic correction (weak localization), an exponential decay (strong localization), or by algebraic localization of the ZEMs. As detailed in the Supplemental Material, the quantum correction to the conductivity [$\delta\sigma(\lambda)=\sigma(\lambda)-\sigma_{sc}$] at $E$=0.4 eV is numerically found to downscale as $\delta\sigma(\lambda)\sim -2e^{2}/(\pi h)\ln(\lambda/\lambda_{e})$ [with $\lambda\equiv \sqrt{\Delta X^{2}(t)}$ the time-dependent wave packet space extension and $\lambda_{e}$ related to $\ell_e$ \cite{LHE_PRB86}]. Differently, for $E$=0.2 eV the length-dependent conductivity exhibits an exponential behavior $\sigma\sim\exp(-\lambda/\xi)$ ($\xi$ the localization length), evidencing a strong-localization regime \cite{EVE_RMP80}. Exactly at the Dirac point, the conductivity decays following a power law $\sigma\sim \lambda^{-2}$. This behavior is actually in full agreement with the localization of the ZEMs measured experimentally by scanning tunneling microscopy \cite{UGE_PRL104}.

\begin{figure}[tpb]
  \begin{center}
  \resizebox{8cm}{!}{\includegraphics{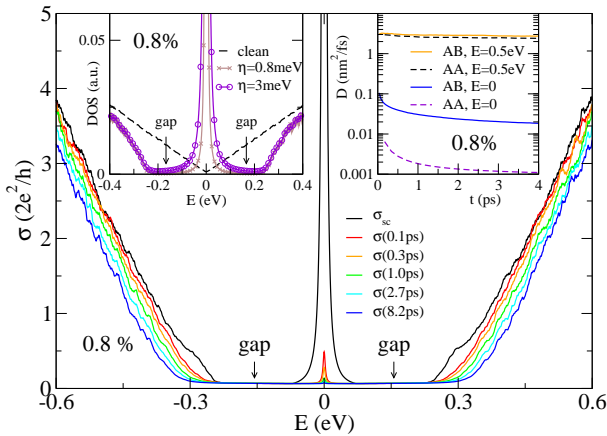}}
  \caption{Main frame: $\sigma_{sc}(E)$ and $\sigma(E,t)$ for graphene (uncompensated case) and energy resolution $\eta=3$ meV. Left inset: DOS with an energy gap revealed by $\eta$ scaling \cite{Note1} and ZEMs. Right inset: Diffusion coefficients at $E=0.5$ eV and $E=0$ ($\eta=3$ meV) for both compensated (AB) and uncompensated (AA) cases. All data for $n=0.8\%$}
	\label{figAA}
	\end{center}
\end{figure}

A remarkably different picture emerges in the uncompensated case, for which the sublattice symmetry is fully broken. The DOS shown in Fig.\ref{figAA} (left inset) evidences the presence of ZEMs sharply peaked at $E=0$. In contrast to the compensated case, the depletion of the low-energy conductivity is here inherited from the presence of energy gaps \cite{PER_PRL96,PER_PRB77}. The semiclassical conductivity strongly increases when approaching the Dirac point, much more than in the compensated case and also increases when improving the energy resolution.  However, the large value of $\sigma_{sc}$  does not reflect the extendedness of the corresponding ZEMs. This can be rationalized by scrutinizing $\sigma(E=0,t)$ and $D(E=0,t)$, which are actually strongly decaying with time. Indeed $D(E=0,t)$ becomes extremely small compared to that at finite energies (e.g. at 0.5 eV) and much smaller compared to the compensated case with the same vacancy concentration (see Fig.\ref{figAA}, right inset). Additionally, $D(E=0,t)$ decays when improving the energy resolution (not shown here), thus demonstrating that although many ZEMs are present, they do not participate in conduction, and that the large value of $\sigma_{sc}$ obtained numerically results from the high DOS at $E=0$. 
Furthermore, the physical relevance of a semiclassical conductivity at the Dirac point is highly questionable (see also Supplemental Material). 
For the quantum conductivity, on the other hand, the strong decay of $\sigma(E=0,t)$ with time is consistent with localized modes similar to the compensated case.
We also find that away from the Dirac point a higher energy resolution reduces $\sigma_{\text sc}$ and $\sigma(t)$ as observed for the DOS, thus unambiguously indicating the energy gap as the origin of the conductivity decrease, and ruling out any diffusive regime and Anderson localization phenomenon.
Finally, for larger energies away from the gap region, one observes that the wave packet dynamics for the compensated (AB) and uncompensated (AA) cases are very similar; see Fig.\ref{figAA} (right inset). This discards any singular transport mechanism in an uncompensated situation, which is different from previous reports on hydrogenated graphene \cite{LEC_ACS5}.

\begin{figure}[tpb]
  \begin{center}
	\resizebox{8cm}{!}{\includegraphics{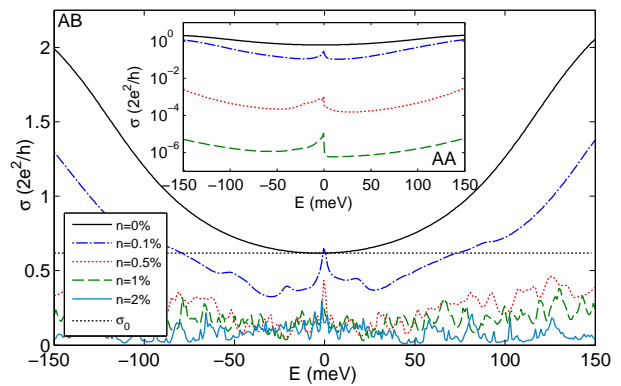}}
	\caption{Main frame: Conductivity for strips with $W=150$ nm, $L=15$ nm, and a compensated vacancy density up to $2\%$. Inset: Same information for uncompensated vacancies with densities up to $1\%$.	}
	\label{landauer_uncompensated}
	\end{center}
\end{figure}

\textit{ZEMs effects in disordered finite graphene strips}.- 
In contrast to two-dimensional graphene, the role played by ZEMs in transport through finite strips in between highly doped contacts turns out to be quite different.  In this configuration, the contacts have much a higher density of propagating states than the central strip, especially at the Dirac point. Accordingly, many states from contacts tunnel through the strip as evanescent modes, yielding a minimum ballistic value $\sigma_{\rm min}=4e^{2}/\pi h$ for clean samples \cite{TWO_PRL96,KAT_EPJB51,MIA_SCI317}. The presence of ZEMs increases the number of available states at the Dirac point in the central strip. Two competing transport mechanisms then drive the conductivity behavior, namely an enhanced tunneling probability assisted by ZEMs together with multiple scattering and quantum interferences, which develop owing to the randomness of vacancies distribution.  

\begin{figure}[tpb]
	\begin{center}
	\resizebox{8cm}{!}{\includegraphics{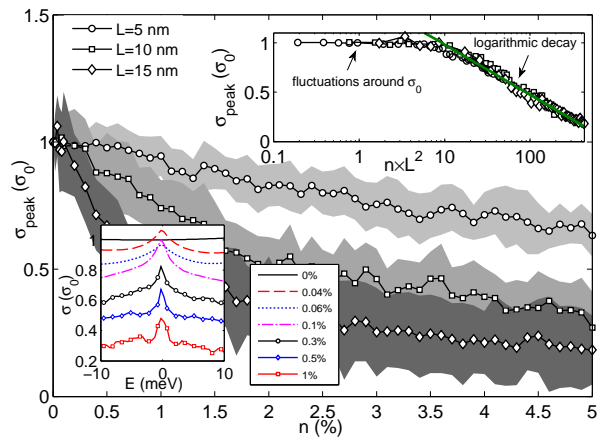}}
	\caption{Main frame: Average conductivity peak versus $n$ for strips with $W = 150$ nm and $L = 5$, 10, and 15 nm. The shaded areas around the curves indicate the standard deviation with respect to the average value. Top inset: Same as the main frame but as a function of $n\times L^2$. The thick straight line is a guide to the eye. Bottom inset: Average conductivity for $W = 150$ nm, $L = 15$ nm, and various $n$.
	}
\label{landauer_compensated}
\end{center}
\end{figure}

Figure \ref{landauer_uncompensated} (main frame) shows the quantum conductivity $\sigma$ for a strip with length $L=15$ nm, width $W=150$ nm, and compensated vacancy density in the range $[0\%,2\%]$. In the absence of vacancies, $\sigma$ shows the minimum conductivity $\sigma(E=0)\equiv \sigma_0\approx \sigma_{\rm min}$ expected for the ballistic limit when $L\ll W$ (see the horizontal dotted line) \cite{TWO_PRL96}. For $n=0.1\%$, the strip length is close to the mean-free path; see Fig.\ref{figAB}. Therefore, the transport along the strip remains quasiballistic, a fact further confirmed by the smooth decay of $\sigma$ all over the spectrum except at the Dirac point, where $\sigma$ keeps a larger value. For higher densities and away from the Dirac point, the decay of $\sigma(E)$ with $n$ is consistent first with the occurrence of a diffusive regime and then with localization phenomena, as revealed by the strongly fluctuating conductivity. Note that despite the few nanometers short mean-free path, even for $n=2\%$, the conductivity remains significant as a consequence of the large number of conductive channels that penetrate the undoped strip.
The conductivity around the Dirac point is further scrutinized in Fig.\ref{landauer_compensated} (bottom inset) for strips with $L=15$ nm,  $W=150$ nm, and compensated vacancy densities up to $n=1\%$. To reduce sample-to-sample fluctuations, all the results were averaged over 20 random disordered configurations. Far from the Dirac point, the conductivity is found to decrease regularly with $n$. At $E=0$, notably enough, a peak is always present, which can slightly exceed $\sigma_0$ at very low density ($n\lesssim 0.04\%$).
This indicates that the ZEMs generated at the Dirac point are sufficiently delocalized to assist (and even enhance) electron tunneling through the strip. Backscattering becomes eventually dominant for sufficiently high defect concentration, as manifested by the smooth conductivity decrease.  The dependence of the conductivity peak ($\sigma_{\rm peak}$) on the different system parameters is reported in Fig.\ref{landauer_compensated} (main frame) for compensated vacancy densities up to 5\% and lengths up to 15 nm. The decrease of $\sigma_{\rm peak}$ with $n$ is very slow, especially for the shortest strip, and even for strong disorder ($n=5\%$) $\sigma_{\rm peak}$ remains significantly large.
As illustrated in Fig.\ref{landauer_compensated} (top inset), $\sigma_{\rm peak}$ is actually a universal function of $n\times L^{2}$. Remarkably enough, $\sigma_{\rm peak}$ fluctuates around or goes slightly above $\sigma_{0}$ for very low $n\times L^2\lesssim 10$, thus supporting the possibility for a supermetallic state, introduced by Ostrovsky and co-workers \cite{OST_PRB74,OST_PRL105}. For $n\times L^2\gtrsim 10$, $\sigma_{\rm peak}$ decreases roughly logarithmically, as the result of finite size effects and proximity between vacancies.

The conductivity of graphene strips (with $W=150$ nm, $L=$15 nm, and $n$ up to 1\%) for uncompensated vacancies is reported in Fig.\ref{landauer_uncompensated} (inset). In marked contrast with the prior case, a gap develops at low density together with a reduced but finite conductivity peak at $E=0$.  As for the case of two-dimensional graphene (Fig.\ref{figAA}), the gap formation leads to the suppression of tunneling due to the almost vanishing DOS. The Dirac conductivity peak is a signature of the highly localized nature of zero-energy states generated by uncompensated vacancies \cite{PER_PRB77}, which are not enough spatially extended to significantly contribute to tunneling and obviate to the DOS decrease.
More details on the energy gap scaling and, in general, on the transport properties away from the Dirac point will be published elsewhere \cite{CRE_CRY3}. 

In conclusion, the contribution of ZEMs to quantum transport in disordered graphene has been discussed for various transport geometries and sublattice symmetry-breaking situations. 
Our findings provide a broad overview of the low-energy transport phenomena in graphene in the presence of ZEMs, including the formation of an insulating state at the Dirac point, accessible in absence of electron-hole puddles \cite{PON_NP7}. The role of electron-electron interaction (here neglected), might also play some important role in capturing the full picture and deserves further investigation \cite{VOZ_PS146,ELI_NP7}.

S.R. acknowledges the Spanish Ministry of Economy and Competitiveness for national project funding (MAT2012-33911), and SAMSUNG for support within the Global Innovation Program.

\clearpage
\newpage
\setcounter{page}{1}
\setcounter{figure}{0}
\onecolumngrid
\section*{Supplemental Material\\Broken Symmetries, Zero-Energy Modes and Quantum Transport in Disordered Graphene: From Supermetallic to Insulating Regimes}
\subsection*{by Alessandro Cresti, Frank Ortmann, Thibaud Louvet, Dinh Van Tuan, and Stephan Roche}
\twocolumngrid

\begin{figure}[t]
	\begin{center}
	\resizebox{8cm}{!} {\includegraphics{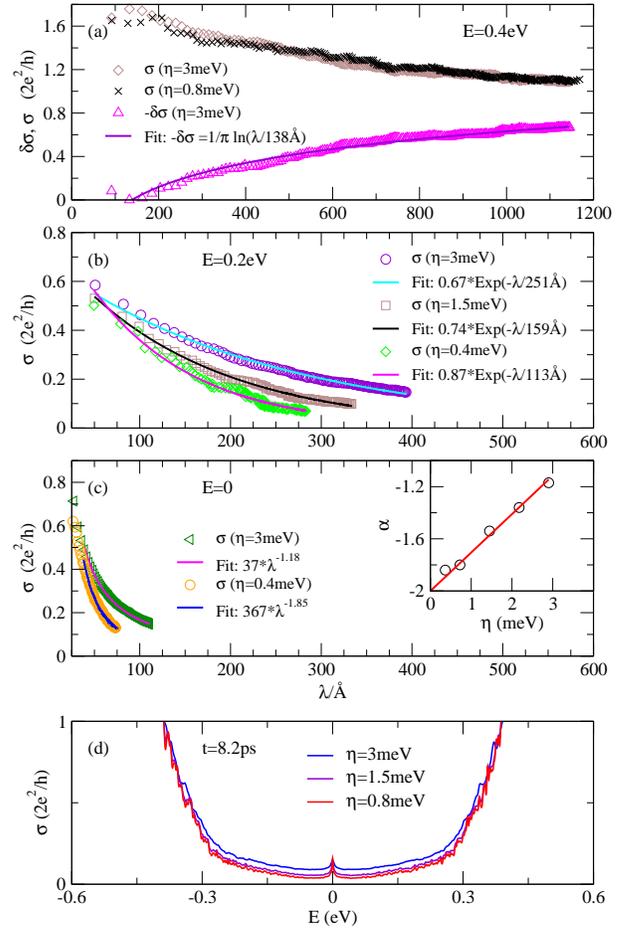}}
	\renewcommand{\figurename}{FIG. S$\!\!$}
	\caption{Length-dependent conductivity for different energies and $0.8\%$ vacancy concentration in the compensated case. 
	(a) Conductivity $\sigma$ and quantum correction $\delta\sigma=\sigma-\sigma_{sc}$ at $E=0.4$ eV. The logarithmic fit confirms the weak-localization regime. 
	(b) Low energy conductivity ($E=0.2$ eV) and corresponding fit indicate Anderson localization regime. 
	(c) At zero energy the conductivity decay is even stronger and cannot be fitted with an exponential decay. 
	(d) Conductivity at largest simulated times (8.2ps) and its residual dependence on $\eta$.}
\label{FigS1}
  \end{center}
\end{figure}

In this Supplemental Material, we demonstrate that the decay of the quantum conductivity shown in Fig. 1 (main frame) of the main paper for compensated vacancy concentration of 0.8$\%$ (at finite energies) follows the conventional scaling theory of localization for two-dimensional disordered systems.
To this end, Fig. S\ref{FigS1} shows the conductivity as a function of the wave-packet spread $\lambda(t)=\sqrt{\Delta X^2(t)}$ at different energies and for different energy resolution parameters $\eta$. 

In Fig. S\ref{FigS1} (a), we see that the decay of the conductivity with $\lambda$ observed at $E=0.4$ eV (due to the quantum correction) can be fitted by $-1/\pi\ln(\lambda/\lambda_{e})$ (using units $2e^2/h$ and with $\lambda_{e}$ related to the mean free path \cite{LHE_PRB86s}). This clearly indicates that the system is in the weak-localization transport regime. In the same panel (a) a strongly reduced broadening $\eta=0.8$ meV yields the same conductivity.

At lower energy ($E=0.2$ eV), a strong localization regime is obtained as seen in Fig. S\ref{FigS1} (b), for different energy resolutions from $\eta=3$ meV down to $\eta=0.4$ meV. 
The length-dependent conductivity decays exponentially as $\sigma\sim\exp(-\lambda/\xi)$, hence evidencing the strong-localization regime ($\xi$ the localization length).
Note that this scaling law is observed independently of the energy precision parameter $\eta$, thus indicating that our approach is able to unambiguously catch the physics of the system and that there is only a residual quantitative, but not qualitative, dependence on $\eta$. 
Moreover, the localization length varies only weakly at lowest $\eta$ indicating a 
limit $\xi\approx$10 nm when $\eta\rightarrow 0$, which confirms the reliability 
of the numerical simulation.

At the Dirac point ($E=0$), localization is observed in Fig. S\ref{FigS1} (c) since the conductivity decays with length $\lambda$. However, in contrast to finite energies, it follows a power-law behavior $\sigma\propto \lambda^{\alpha}$ with $\alpha<0$. The inset shows $\alpha$ upon decreasing the broadening $\eta$ down to about 0.4 meV, which is the present limit of our numerical resolution. Note that the observed behavior is consistent with the limit $\alpha=-2$ for $\eta\to 0$, which has been observed experimentally \cite{UGE_PRL104s} for the localization of ZEMs by means of scanning tunneling spectroscopy. The localization at $E=0$ is even stronger than in the Anderson regime and can therefore not be attributed to multiple scattering and quantum interference effects, i.e. the strong localization regime, but is rather a signature of zero-energy modes.
This is further corroborated by the length $\lambda\sim 5$ nm over which $\sigma$ localizes, which is on the same order as the spatial extension of the bound states experimentally measured \cite{UGE_PRL104s}.
 
We would like to point out here that our results for compensated vacancies are well-defined and converge in the limit of small $\eta$. Figure S\ref{FigS1} (d) finally shows that at the largest time considered for the calculation of the conductivity (8.2ps), $\sigma(E)$ is well controlled when decreasing $\eta$, with a more pronounced noise level at smaller $\eta$, an effect which defines a lower limit for $\eta$ to avoid non-physical mathematical singularities.

The case of uncompensated vacancies is more subtle and the role of $\eta$ depends on the region of the energy spectrum considered. At high energies, as seen in Fig. 2 of the main paper, the physics is the same as for compensated vacancies and the limit $\eta\rightarrow 0$ leads to quantitatively robust results in the weak-localization regime. At lower energies, but not at the Dirac point, the system exhibits a genuine energy gap where electrons cannot propagate, thus leading to vanishing conductivity, which is therefore not related to a transition from diffusive to localized regime. The conductivity suppression is therefore of different nature as compared to the compensated case. 

At the Dirac point, our numerical approach provides an $\eta$-dependent value of the quantum conductivity that progressively decreases with $\eta$, consistently with vanishing conductivity. We find that, similarly to Fig. S\ref{FigS1} (c), the conductivity decays on a very short length scale, which is even shorter than for the compensated case. This is observed for all the $\eta$ values used in the simulations (down to 0.4 meV) and with a steeper decay for smaller $\eta$, which is consistent with bound states of vanishing conductivity similar to the compensated case.
In contrast, we observe that the semiclassical conductivity diverges with small $\eta$.
The reason is that, for the uncompensated case, all vacancy-induced modes are exactly at $E=0$ and their corresponding DOS and semiclassical conductivity have a $\delta$-like distribution centered in the gap where no propagation is possible. However, the broadening and the height of the DOS peak (as well as $\sigma_{\text sc}$ peak) are artificially driven by the finite parameter $\eta$.

\end{document}